\documentclass[useAMS,usenatbib,twocolumn,tighten]{aastex61}
\usepackage{graphicx}
\usepackage{times}
\usepackage{amssymb}
\usepackage{amsmath}
\usepackage{sansmath}
\usepackage{aas_macros}
\usepackage{multirow}
\begin{document}

% --- title --- %
\shorttitle{Growth rate constraints from LSST SNe~Ia}
\shortauthors{Howlett et al.}
\title{Measuring the growth rate of structure with Type IA Supernovae from LSST}

\correspondingauthor{Cullan Howlett}
\email{cullan.howlett@icrar.org}

\author{Cullan Howlett}
\affiliation{International Centre for Radio Astronomy Research, The University of Western Australia, Crawley, WA 6009, Australia.}
\affiliation{ARC Centre of Excellence for All-sky Astrophysics (CAASTRO).}

\author{Aaron S.G. Robotham}
\affiliation{International Centre for Radio Astronomy Research, The University of Western Australia, Crawley, WA 6009, Australia.}

\author{Claudia D.P. Lagos}
\affiliation{International Centre for Radio Astronomy Research, The University of Western Australia, Crawley, WA 6009, Australia.}
\affiliation{ARC Centre of Excellence for All-sky Astrophysics (CAASTRO).}

\author{Alex G. Kim}
\affiliation{Lawrence Berkeley National Laboratory, 1 Cyclotron Road, Berkeley, CA 94720, USA.}


\begin{abstract}
We investigate measuring the peculiar motions of galaxies up to $z=0.5$ using Type Ia supernovae (SNe~Ia) from LSST, and predict the subsequent constraints on the growth rate of structure. We consider two cases. Our first is based on measurements of the volumetric SNe~Ia rate and assumes we can obtain spectroscopic redshifts and light curves for varying fractions of objects that are detected pre-peak luminosity by LSST (some of which may be obtained by LSST itself and others which would require additional follow-up). We find that these measurements could produce growth rate constraints at $z<0.5$ that significantly outperform those using Redshift Space Distortions (RSD) with DESI or 4MOST, even though there are $\sim4\times$ fewer objects. For our second case, we use semi-analytic simulations and a prescription for the SNe~Ia rate as a function of stellar mass and star formation rate to predict the number of LSST SNe~IA whose host redshifts may already have been obtained with the Taipan+WALLABY surveys, or with a future multi-object spectroscopic survey. We find $\sim 18,000$ and $\sim 160,000$ SN~Ia with host redshifts for these cases respectively. Whilst this is only a fraction of the total LSST-detected SNe~Ia, they could be used to significantly augment and improve the growth rate constraints compared to only RSD. Ultimately, we find that combining LSST SNe~Ia with large numbers of galaxy redshifts will provide the most powerful probe of large scale gravity in the $z<0.5$ regime over the coming decades.
\end{abstract}

\section{Introduction}
A key science driver of future surveys such as DESI \citep{Levi2013,DESI2016} and 4MOST \citep{deJong2012} is to test General Relativity (GR; \citealt{Einstein1916}). Whilst our consensus cosmological model ($\Lambda$CDM) has strong support from a variety of probes (\citealt{Planck2016}, \citealt{Alam2016}, \citealt{Riess2016}, \citealt{Hildebrandt2017}), the nature of dark energy and matter remains unknown, and tensions exist between these results. Modifying the large scale behaviour of gravity is a promising alternative towards resolving this.

The peculiar velocities (PVs) of galaxies present a method to test gravity. The PV of a galaxy towards an overdensity at scale factor $a$, is dictated by the growth rate of structure, $f(a)=d\,\mathrm{ln}\,D(a)/d\,\mathrm{ln}\,a$, the logarithmic derivative of the linear growth factor $D$. The linear growth factor in turn describes how density perturbations in the Universe grow over cosmological time under the influence of gravity. Within the framework of $\Lambda$CDM and GR the linear growth factor is given by \citep{Heath1977}
\begin{equation}
D(a) = \frac{5}{2}a^{3}\Omega_{m}(a)E^{3}(a)\int^{a}_{0}\frac{da'}{(a'E(a'))^{3}},
\label{eq:growthfactor}
\end{equation} 
where
\begin{align}
\Omega_{m}(a) &= \frac{\Omega_{m,0}}{a^{3}E^{2}(a)}, \\
E(a) &= \sqrt{\frac{\Omega_{m,0}}{a^{3}}+\Omega_{\Lambda,0} + \frac{(1-\Omega_{m,0}-\Omega_{\Lambda,0})}{a^{2}}},
\end{align}
and $H_{0}$, $\Omega_{m,0}$ and $\Omega_{\Lambda,0}$ describe the cosmological model. In turn, GR predicts a scale-independent growth rate that can be approximated as $f(a)\approx\Omega_{m}(a)^{0.55}$ \citep{Linder2007}. Measuring a growth rate that differs from this could be used to falsify GR and constrain alternative theories of gravity. 

Redshift Space Distortions (RSD; \citealt{Kaiser1987}) in the clustering of galaxies are the most commonly used method for constraining the growth rate and the ability to make precise RSD measurements is an integral part of the design of DESI and 4MOST. However this approach is fundamentally limited due to cosmic variance and the degeneracy between $f(a)$ and galaxy bias. 

Direct measurements of PVs can instead be obtained by comparing the redshift-inferred distance to that measured using the intrinsic properties of the galaxy or its inhabitants. Examples include the Tully-Fisher (TF; \citealt{Tully1977}), and Fundamental Plane (FP; \citealt{Dressler1987,Djorgovski1987}) relationships and the use of Type Ia supernovae (SNe~Ia; \citealt{Phillips1993}). These measurements are not affected by galaxy bias \citep{Zheng2015}, probe larger scales than the density field, and can be used to overcome the cosmic variance limit \citep{Park2000,Burkey2004}. \cite{Koda2014} and \cite{Howlett2017} showed that imminent redshift and peculiar velocity surveys, such as Taipan \citep{daCunha2017} and WALLABY \citep{Koribalski2012} have the ability to produce some of the most accurate measurements of the growth rate to date. 

In this work, we consider the capabilities of PV's measured using next generation measurements of SNe~IA. \cite{Bhattacharya2011} and \cite{Odderskov2017} demonstrated that PVs obtained from the large number of SNe~Ia we will detect with LSST have the potential to constrain dark energy and the linear matter variance in spheres of radius $8h^{-1}\,\mathrm{Mpc}$, $\sigma_{8}$. We instead build on the work of \cite{Howlett2017} to show that, given host galaxy redshifts and accurate SNe classification, the two-point correlations between the velocities and positions of these SNe~Ia present a unique opportunity to measure the growth rate in the $z<0.5$ universe. Using Fisher matrix forecasts, we find that these measurements could significantly improve over the constraints using just RSD with DESI and 4MOST. 

Our aim is to present the constraints possible with SNe~IA that will be \textit{detected} (pre-peak luminosity) with LSST. However, LSST itself will only measure accurate light curves for a small percentage of these within its wide field survey. Additional follow-up will be needed to obtain host redshifts and spectroscopic classifications for all SNe~IA, and improve on the overall photometric data quality and volume. Hence we provide forecasts for a variety of scenarios ranging from the typical numbers of SNe~IA that may have accurate light curves from LSST itself, to the case where we can use additional follow-up to obtain light curves for all LSST detections. We then investigate the LSST-detected SNe~IA we could expect to also have host redshifts from upcoming large galaxy surveys. Through this, we seek to motivate further consideration of the overlap between LSST and future spectroscopic surveys, the need for accurate photometric or spectroscopic follow-up of SNe~IA whose light curves or types cannot be measured by LSST alone, and studies into how well PVs could be measured with SNe~IA given realistic simulations of LSST.

Throughout, we quote AB magnitudes and assume a cosmology of $\Omega_{m}=0.3121$, $\Omega_{b}=0.0488$, $H_{0}=100h=67.51\mathrm{km\,s^{-1}\,Mpc^{-1}}$, $n_{s}=0.9653$ and $\sigma_{8}(z=0)=0.815$. 
\newpage
\section{Peculiar Velocities with LSST SNe~Ia} \label{sec:LSST}
The Large Synoptic Survey Telescope (LSST; \citealt{Ivezic2008}) project is a planned photometric survey whose large field-of-view and high cadence will allow for high resolution imaging of approximately half the sky to be taken every few days. These properties will allow for the detections of millions of SNe~Ia over the course of the survey. Measurements of the velocity field can be obtained from such a sample of SNe~IA by taking the difference between the SNe~Ia absolute magnitudes measured from their light curves and inferred from their apparent magnitudes and host galaxy redshifts \citep{Johnson2014, Huterer2016}. Equivalently, given a measurement of the  distance modulus $\mu$, and the host redshift $z$, we can define the `log-distance' ratio $\Delta d$, the logarithm of the ratio between the comoving distance inferred from the redshift $d_{z}$ (in parsecs), and the true comoving distance,
\begin{equation}
\Delta d = \mathrm{log}_{10}\left(\frac{d_{z}}{10\mathrm{pc}}\right) - \frac{\mu}{5}
\end{equation}
The log-distance ratio can then be related to the peculiar velocity using, for example, the estimator of \cite{Watkins2015}
\begin{equation}
v \approx \frac{cz_{m}}{1+z_{m}}\mathrm{ln}(10)\Delta d
\end{equation}
where $z_{m} = z[1 + 1/2(1-q_{0})z - 1/6(1-q_{0}-3q_{0}^{2}+j_{0})z^{2}]$ and $q_{0}$ and $j_{0}$ are the deceleration and jerk parameters. With the same sample of host galaxy redshifts we can also consider measurements of the density field and cross-correlations between the density and velocity fields.

In addition to the large numbers of measured PVs from LSST-detected SNe~IA, the smaller intrinsic scatter in the SN~Ia distance relationship compared to the TF or FP relations makes each one more useful for constraining gravity. We do not expect to be able to reduce the intrinsic scatter for TF or FP galaxies below $20\%$ for even next generation surveys, but the distance error for SNe~Ia is currently at the $10\%$ level \citep{Rest2014} and could be reduced to as little as $5\%$ in the coming decades \citep{Fakhouri2015}. This allows us to probe the velocity field on larger scales and at higher redshifts than is currently possible. We focus on measurements of the two point correlations between the density and velocity fields that will be obtainable with LSST SNe-Ia, although other statistics, such as our local `bulk flow', could also be measured. 

For all numbers in this work we assume a ten-year LSST survey, and sky coverage of $18,000\,\mathrm{deg^{2}}$. The LSST survey design we adopt for our forecasts is based on the LSST Observing Strategy White Paper (LSST Science Collaborations, in preparation)\footnote{This is a ``living document", so more specifically we use Version 0.99.d28199b found online at \url{https://github.com/LSSTScienceCollaborations/ObservingStrategy/tree/master/whitepaper}.}. Simulations of the LSST observing strategy suggest $\sim40\%$ of $z<0.5$ SNe~IA will be detected pre-peak luminosity and hence suitable for lightcurve measurements. When discussing LSST-detected SNe~IA, we are presenting numbers and forecasts weighted by this, i.e., we multiply the volumetric rate and SN~IA rate as a function of stellar mass and star formation rate in the remainder of this work by $0.4$.

\section{LSST SNe~Ia Numbers}
\subsection{Volumetric rate} \label{sec:volrate}
To predict the numbers of SNe~Ia with PV measurements, we consider two different scenarios. First, we take a measurement of the volumetric rate of $6.8\times10^{-5}(1+z)^{2.04}h^{3}\,\mathrm{SN}$~$\mathrm{Ia\,yr^{-1}\,Mpc^{-3}}$ \citep{Dilday2010}. For our adopted LSST survey this gives a total of $120\,\mathrm{SN}$~$\mathrm{Ia\,deg^{-2}}$ up to $z=0.5$ and $\sim2.2\times10^{6}\,\mathrm{SN}$~$\mathrm{Ia}$ in total. Of the SNe~IA that LSST will detect, only a small fraction of those in the wide field survey will have enough repeat visits for accurate light curves to be measured. The latest simulations from the LSST Observing Strategy White Paper predict on the order of $\sim 50,000\,\mathrm{SN}$~$\mathrm{Ia\,yr^{-1}}$ with accurate LSST light curves may be achievable for certain observing strategies. In the interest of motivating follow-up from other instruments, we consider forecasts for different numbers of LSST-\textit{detected} SNe~IA between those $\sim 50,000\,\mathrm{yr^{-1}}$ that LSST may obtain distances for, up to the full number of $\sim 220,000\,\mathrm{yr^{-1}}$.

In all cases (including the following section), we assume that the LSST (and follow-up) observing strategy is designed so that the SNe~IA with good light curves are randomly distributed within its wide footprint and that host redshifts and spectroscopic classifications can be obtained for these. In Section~\ref{sec:systematics} we will discuss how the requirements for spectroscopic classification could be relaxed given accurate photometric typing, and the possible impact of systematic errors this could introduce. However, this is still an active area of research and the accuracy of photometric classifiers in the era of LSST is largely unknown, so for the purposes of our forecasts perfect classification is assumed. We will also investigate SN~IA that may already have host redshifts from large galaxy surveys, which it is logical to prioritise for follow up. Whilst obtaining light curves, host redshifts and spectroscopic classifications for all LSST-detected SNe-IA is optimistic, the number of galaxy redshifts is far below the number of targets observable with next generation spectroscopic instruments and we expect nearly $100\%$ of spectroscopic targets below $z=0.2$ (which have the most accurate distances) to be `cheap' to obtain with a $1-2$m telescope\footnote{For a comparison of current and future surveys see \url{http://compare.icrar.org/}}.

\subsection{Pre-selected SNe~Ia hosts} \label{sec:surveys}
For the second scenario we consider LSST-detected SNe~Ia for which host redshifts may have already been obtained by large spectroscopic galaxy surveys prior to or during LSST operations. These SNe~Ia can be used in addition to the full spectroscopic galaxy sample to improve over the constraints from RSD alone. As such, these targets are the logical choice for additional follow-up if required, and many may already have LSST distance measurements. This is especially true considering, as we will show, the small number of SNe per year and the constraining power they offer when combined with the planned galaxy redshift surveys. In the following, we consider the combined Taipan \citep{daCunha2017} and WALLABY \citep{Koribalski2012} surveys, and a future spectroscopic sample with target density similar to DESI/4MOST.

To predict the number of SNe~Ia we combine a simulated galaxy catalogue with observationally-constrained models for the SNe~Ia rate as a function of stellar mass and star formation rate (SFR). Our simulated catalogue uses the \cite{Lagos2012} variant of the semi-analytic model \textsc{galform} \citep{Cole2000}, which was run on merger trees constructed by \cite{Jiang2014} from the Millennium N-body simulation \citep{Springel2005} and has an effective stellar mass limit of $10^8M_{\odot}$. Lightcones of $z<0.5$ and $1/16^{\mathrm{th}}$ full-sky area were constructed down to $r<24.0$ using the algorithm described in \cite{Merson2013}. Our additional selection functions are then applied on top of this. These lightcones reproduce the luminosity function and number counts of galaxies quite well from the near-UV to the IR \citep{Lagos2014,GonzalezPerez2014,Lacey2016}. Both semi-analytic models and hydrodynamical simulations typically give SFRs and colours that are up to $50\%$ too low and $0.1$ magnitudes too blue, respectively (e.g. \citealt{Mitchell2016,Lacey2016}). We find that artificially increasing these in the simulation \textit{increases} the number of SNe~Ia by $\sim30\%$, which makes our forecasts conservative.

\begin{figure}
\centering
\includegraphics[width=0.49\textwidth]{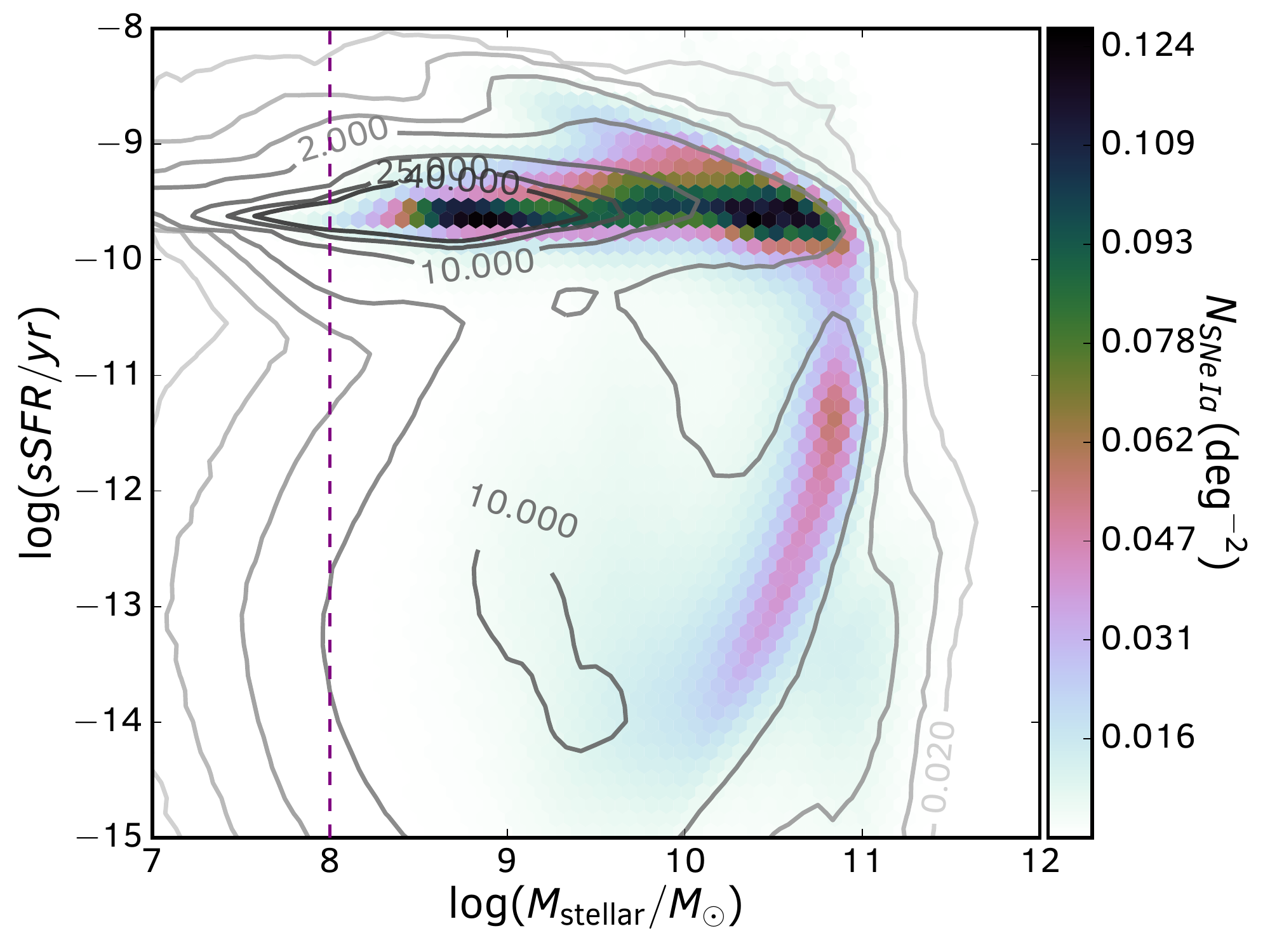}
  \caption{Numbers of SNe~Ia (coloured bins) and galaxies (grey contours) per deg$^{2}$ for our `pre-selected' scenario as a function of stellar mass and specific SFR. The vertical dashed line denotes the effective resolution limit within our simulation.}
  \label{fig:sfr}
\end{figure}

For the expected number of SNe~Ia in these galaxies we use Eq. 5 from \cite{Smith2012}. Fig.~\ref{fig:sfr} shows the number of LSST-detected SNe~Ia and the total number of galaxies in our simulation as a function of stellar mass and specific SFR. As explained in \cite{Smith2012}, large, late-type galaxies are the dominant source of SNe~Ia. Massive, passive galaxies are relatively inefficient producers of low redshift supernovae due to their old stellar populations, whilst the SFR tends to evolve slowly with redshift, such that galaxies with a high current SFR are likely to have had a high SFR in the past, giving rise to the majority of SNe~Ia below $z=0.5$.

The total number of $z<0.5$ SNe~Ia from the simulation, $29.3\,\mathrm{SN}$~$\mathrm{Ia\,deg^{-2}}$, is a factor of $\sim4$ lower than from the volumetric rate in Section~\ref{sec:volrate}. This discrepancy stems from the different methods for measuring the SNe~Ia rate and inconsistencies between measurements of the SFR and stellar mass densities. For example, \cite{Smith2012} is consistent (depending on the exact model used) with \cite{Dilday2010} if one uses the measured densities of \cite{Hopkins2008} to convert between the two. They are not if we use the densities from our simulation or from more recent studies (Driver et. al., in preparation). However, these inconsistencies reduce the number of SNe~Ia in the simulation relative to the volumetric rate (which is the simpler, and likely more robust measurement) and even in this case, we find that SNe~Ia can be used to significantly augment the growth rate constraints from RSD alone.

\begin{figure*}
\centering
\includegraphics[width=0.49\textwidth]{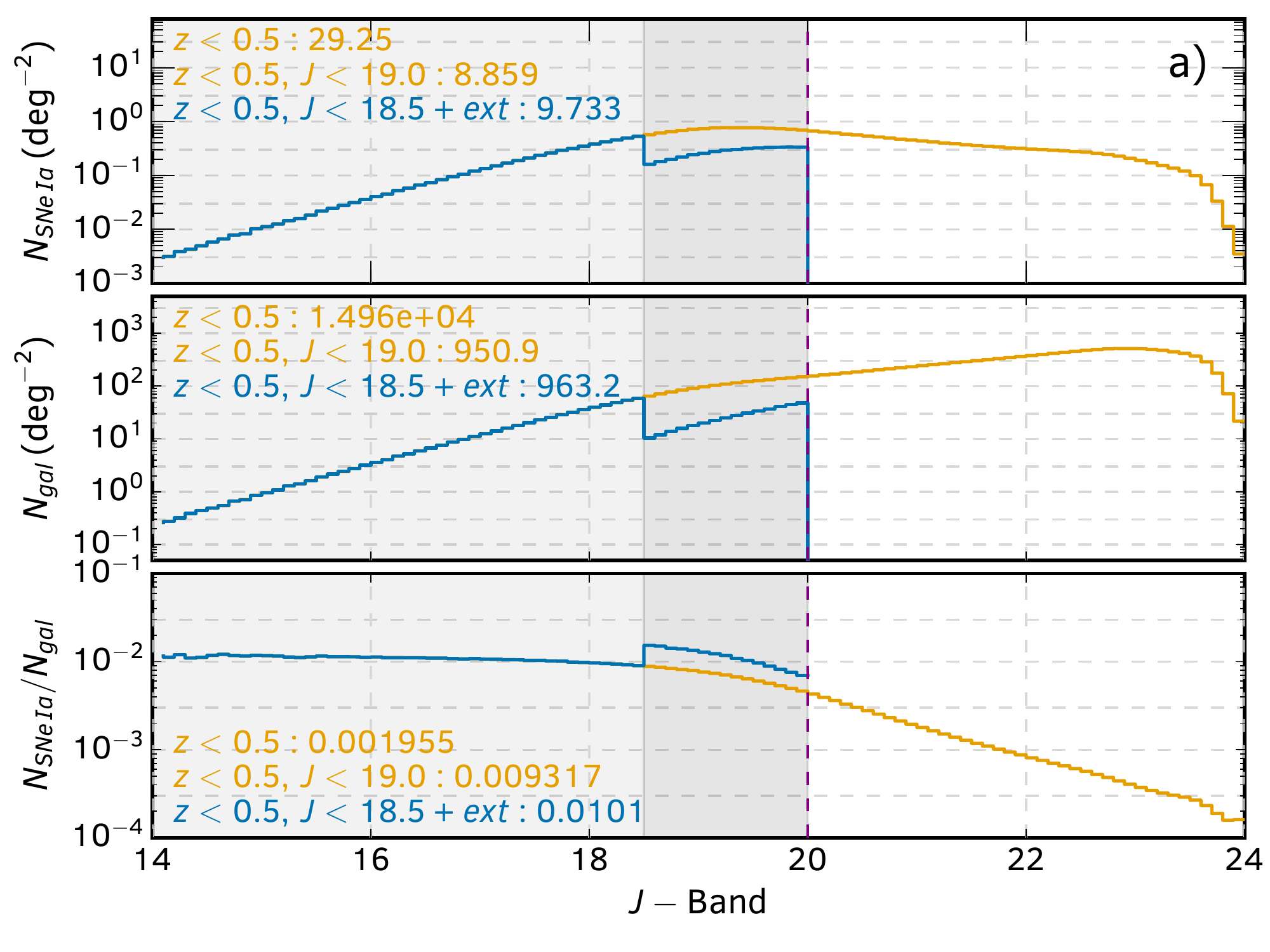}
\includegraphics[width=0.49\textwidth]{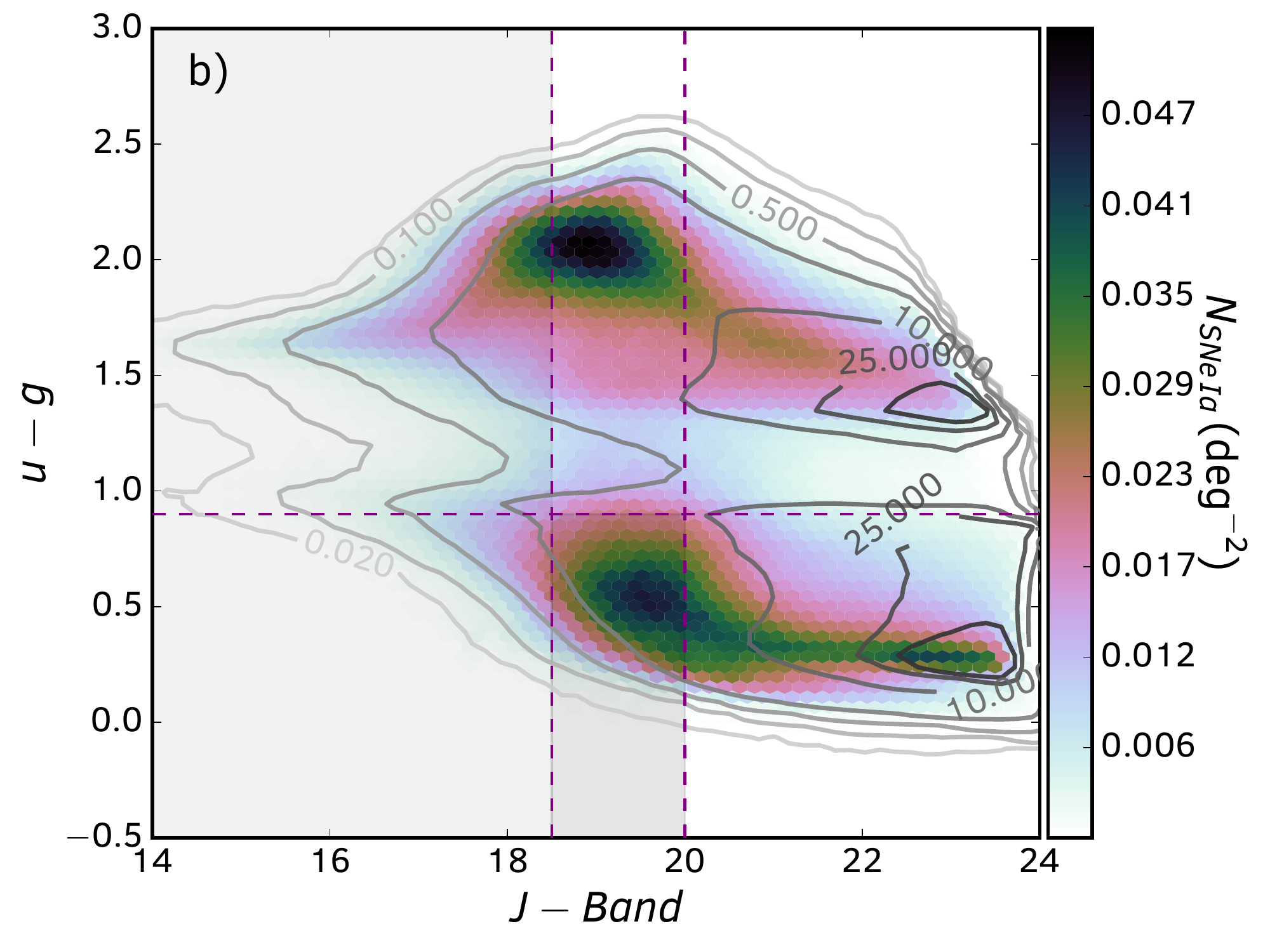}
  \caption{\textbf{a).} Numbers of SNe~Ia, galaxies and the efficiency of SN~Ia production per deg$^{2}$ per bin of 0.1 dex for our `pre-selected' scenario as a function of $J$-band magnitude, for the full catalogue (orange) and our ideal pre-selection (blue). We also give the total numbers for these selections. \textbf{b).} Numbers of SNe~Ia (coloured bins) and galaxies (grey contours) as a function $u-g$ colour and $J$-Band magnitude. In both cases shaded regions indicate the area we choose as our optimal selection: $J<18.5$ plus an extension to $J<20.0$ with $u-g<0.9$.}
  \label{fig:optimal}
\end{figure*}

\begin{figure}
\centering
\includegraphics[width=0.49\textwidth]{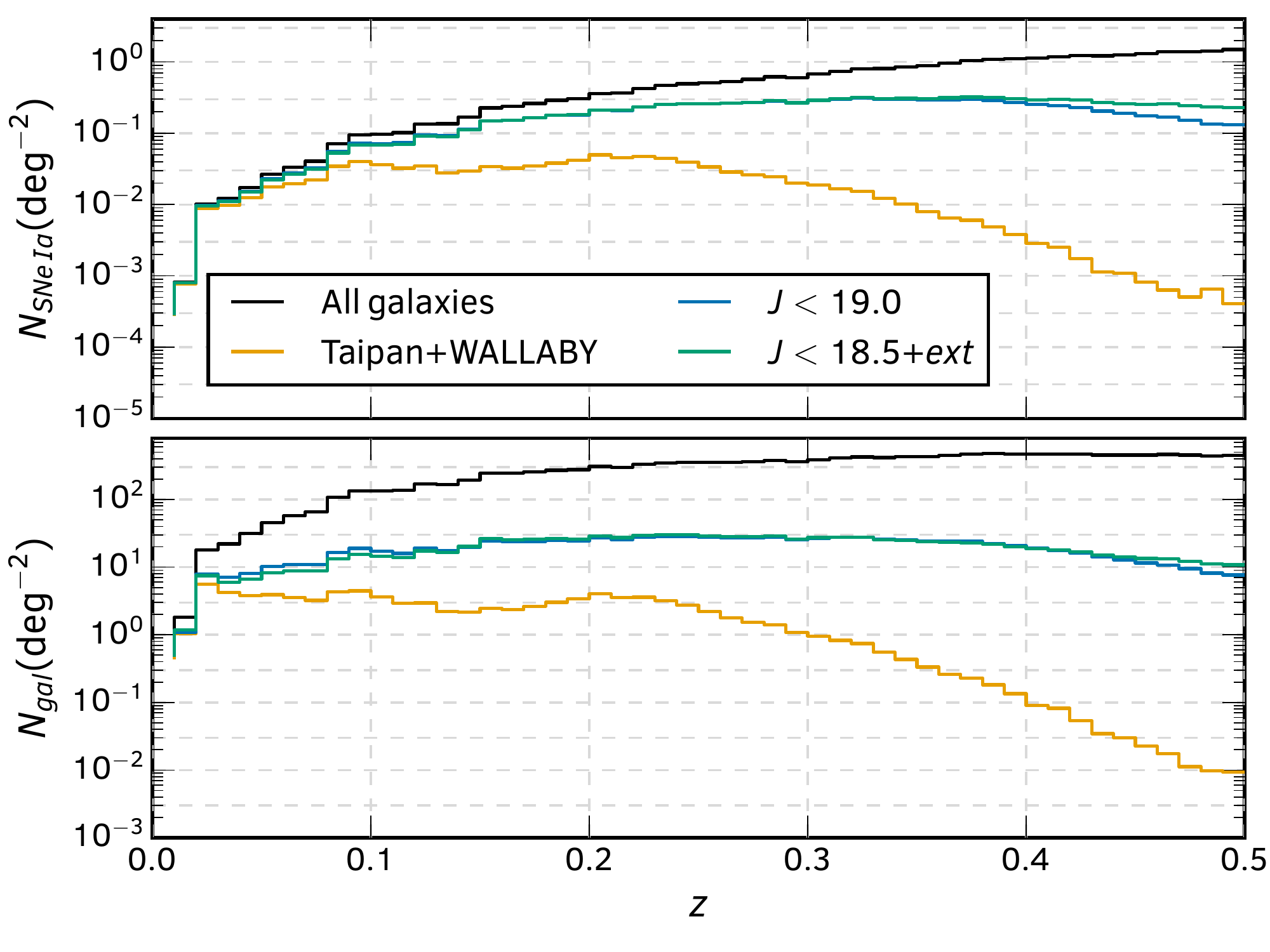}
  \caption{Numbers of SNe~Ia and targets per deg$^{2}$ per $dz=0.01$ bin for our `pre-selected' scenario as a function of redshift for all objects and for the three selections given in Section~\ref{sec:surveys}.
  }
  \label{fig:optimalred}
\end{figure}
\newpage
\subsubsection{Taipan and WALLABY} \label{sec:taipan}
We first consider the number of redshifts we could already have from the near-future Taipan and WALLABY surveys. Starting in 2017, the Taipan galaxy survey on the 1.2m UK Schmidt Telescope will obtain optical spectra for over two million $z<0.4$ galaxies across the southern sky ($\delta \lesssim 20$, $|b|\gtrsim 10$). The current design consists of a five year survey and uses 150 spectroscopic fibres (with a proposed upgrade to 300) spread across a 6-degree focal plane. The final dataset will contain both a magnitude limited $i<17.0$ sample and an LRG extension satisfying $17.0<i<18.1$ and $g-i>1.6$, and this is the selection function we apply to our mock catalogue. The sky coverage of Taipan overlaps almost fully with that of LSST and so we expect host redshifts to be already obtained for many SNe~IA whose host galaxies satisfy either of these selection criteria. In addition to this, $\sim 50,000$ of the galaxies Taipan observes will have high enough signal to noise that they can be placed on the Fundamental Plane and used as distance indicators. Prior to the era of 4MOST/DESI and the opportunities presented with LSST-detected SNe~IA, this will be the largest single PV survey, and of particular interest are those sources that will have both FP distances and PVs measured using SNe~IA. Such a sample will allow for a much greater control of systematics in the peculiar velocity measurements from both SN-Ia and the FP relationships.

Complementary to Taipan, the WALLABY survey \citep{Koribalski2012} is a planned 21-cm HI survey using the Australian SKA Pathfinder (ASKAP), which will cover three quarters of the full-sky ($\delta < 30$) up to $z=0.25$. Hence we expect full angular overlap between this survey and LSST. The survey uses newly designed phased array feeds with 30" resolution over a frequency range of 1.13 to 1.43GHz whilst still allowing for a large $30\mathrm{deg}^{2}$ field of view. The nomimal $1\sigma$ noise limit is expected to be $1.592\mathrm{mJy}\,\mathrm{kms^{-1}}$ and in this work we consider all $5\sigma$ sources. WALLABY will be much more sensitive to low redshift star-forming galaxies than Taipan (see Fig. 11 in \citealt{daCunha2017}), which due to their high star formation rate are still relatively efficient producers of SNe~IA, and will measure redshifts to $\sim500,000$ galaxies, many of which will be missed by Taipan. As with Taipan, a significant fraction of these ($\sim 30,000$) are also expected to have peculiar velocity measurements, this time determined via the Tully-Fisher relation which will also be useful for reducing systematics. 

From the selections for Taipan and WALLABY combined we find $1.0\,\mathrm{SN}$~$\mathrm{Ia\,deg^{-2}}$ and $\sim18,000$ galaxies hosting LSST-detected SNe~Ia, assuming a full overlap area of $18,000\,\mathrm{deg^{2}}$. Even accounting for the factor of four difference between our volumetric rate based and simulation based predictions, this is only $\sim7,200$ SNe-IA per year of LSST operation and so well within the expected number that we could obtain with LSST alone, or with minimal follow-up. 

\subsubsection{A future multi-object spectroscopic survey} \label{sec:futuresurvey}
We then see how many hosts could be obtained from a future multi-object spectroscopic survey similar to DESI or 4MOST. These two multi-pass instruments will have $\sim5000$ and $\sim1600$ usable fibres respectively, spread over $7.5\,\mathrm{\deg}^{2}$ and $4.1\,\mathrm{deg}^{2}$ fields-of-view \citep{Levi2013,deJong2012}. We do not tailor our selection to the requirements of any particular survey, but find that a high efficiency (ratio of the number of SNe~Ia per target) is achieved with a selection close to that of the 4MOST Bright Galaxy (BG) sample. For a magnitude limited sample, J-band magnitudes allow for the highest efficiency; a $J<19.0$ limit gives a target density of $951\,\mathrm{deg^{-2}}$, $8.9\,\mathrm{SN}$~$\mathrm{Ia\,deg^{-2}}$, and $0.9\%$ percent of targets contain an LSST-detected SN~Ia. We can slightly increase the efficiency using a $u-g$ colour cut. A sample consisting of $J<18.5$, plus an extension to $J<20.0$ with $u-g<0.9$ gives a similar target density, but increases the SNe~Ia density to $9.7\,\mathrm{SN}$~$\mathrm{Ia\,deg^{-2}}$. For comparison, observing the same target density but distributed randomly below $J<20.0$ gives $7.1\,\mathrm{SN}$~$\mathrm{Ia\,deg^{-2}}$, a decrease in efficiency of $\sim30\%$. Overall, we predict $\sim160,000\,\mathrm{SN}$~$\mathrm{Ia}$ detected by LSST which could have host redshifts from a future $J<19.0$ survey across $18,000\,\mathrm{deg^{2}}$. Again, this number is small enough that a large fraction of such SNe~IA could have light curves measured by LSST, although follow-up will likely be required to reach the full number, and we advocate prioritising these targets that already have host galaxy redshifts.

Our two selections are summarised in Fig.~\ref{fig:optimal}, where we plot the number of targets and SNe~Ia per deg$^{2}$, and the efficiency as a function of J-band magnitude.  
We also show the $u-g$ colour against $J$-band magnitude, highlighting the area we would preferentially target. Beyond our current selection the efficiency begins to fall significantly, hence obtaining additional host redshifts using dedicated programmes may be preferable to a fainter pre-selection on planned large galaxy surveys.

The total number of galaxies and SN-Ia as a function of redshift for all of our selections is shown in Fig.~\ref{fig:optimalred}. The Taipan and Wallaby surveys will measure many host redshifts for low redshift SNe~IA, however this quickly drops off due to the sensitivity of these surveys. For a future multi-object spectroscopic survey, the number of SNe~IA hosts we will obtain redshifts for remains high 
even up to $z=0.5$, as the selections we consider mainly miss fainter or less star-forming galaxies which are less efficient producers of SNe~IA. These numbers of SNe~IA are used as input for our forecasts in the following section.

Unfortunately, there is no current or planned photometry across the full southern hemisphere that could achieve our colour selection. The current best option, the SkyMapper survey \citep{Keller2007} will only go as faint as $u=20.7, g=21.7$. Including these constraints (and re-examining the other photometric bands under similar limits) shows that a complex selection would be required to improve beyond a simple $J<19.0$ sample. Hence, this is the one we present in our forecasts.

\section{Fisher Matrix forecasts on the growth rate} \label{sec:forecasts}
\subsection{Method}

We forecast the constraints on the growth rate using the Fisher matrix method of \cite{Howlett2017}, modelling the information contained in the two-point correlations between the density field measured using the galaxy redshifts and the velocity field from the SN~Ia PVs.  We have updated the \cite{Howlett2017} models to account for the redshift dependence of the power spectra, growth rate and galaxy bias, but otherwise the method remains unchanged. As such, we present only a brief overview here and we refer the reader to \cite{Howlett2017} for a more complete description. The version of the code used to produce the growth rate forecasts in this paper is publicly available at \url{https://github.com/CullanHowlett/PV_fisher}.

For given parameters of interest $\boldsymbol{\lambda}$, we compute the corresponding elements of the Fisher Matrix $\boldsymbol{\mathsf{F}}$, as
\begin{align}
F_{ij} & = \frac{\Omega_{sky}}{4\pi^{2}} \int^{r_{max}}_{r_{min}} r^{2} dr \int^{k_{max}}_{k_{min}} k^{2} dk \int^{1}_{0} d\mu_{\phi} \, \nonumber \\
& \mathrm{Tr}\left[\boldsymbol{\mathsf{C}}^{-1}(r,k,\mu_{\phi})\frac{\partial{\boldsymbol{\mathsf{C}}}(r,k,\mu_{\phi})}{\partial{\lambda_{i}}}\boldsymbol{\mathsf{C}}^{-1}(r,k,\mu_{\phi})\frac{\partial{\boldsymbol{\mathsf{C}}}(r,k,\mu_{\phi})}{\partial{\lambda_{j}}}\right],
\label{eq:Fisherrad}
\end{align}
where $\Omega_{sky}$ is the sky coverage of the survey, $r_{max}$ ($r_{min}$) are the comoving distances corresponding to the upper (lower) redshift limits of each redshift bin, and we set $k_{max}=0.2h\,\mathrm{Mpc^{-1}}$ and $k_{min} = 2\pi/r_{max}$. $\mu_{\phi}$ is the cosine of the angle $\phi$ between the $k$-vector and the observer's line-of-sight.

The covariance matrix, $\boldsymbol{\mathsf{C}}$ consists of the anisotropic density-density, density-velocity and velocity-velocity power spectra $P_{\delta \delta}$, $P_{\delta v}$ and $P_{vv}$ respectively, as well as the noise associated with each of these,
\begin{equation}
\boldsymbol{\mathsf{C}}(r,k,\mu_{\phi}) = 
\begin{bmatrix}
P_{\delta \delta}(r,k,\mu_{\phi}) + \frac{1}{\bar{n}_{\delta}(r)} & P_{\delta v}(r,k,\mu_{\phi}) \\
P_{\delta v}(r,k,\mu_{\phi}) & P_{vv}(r,k,\mu_{\phi}) + \frac{\sigma^{2}_{obs}(r)}{\bar{n}_{v}(r)}
\end{bmatrix}.
\label{eq:cov}
\end{equation}
The shot-noise in these measurements is inversely proportional to the galaxy number density $\bar{n}_{\delta}(r)$ for the density field, and to the average PV error divided by the SN-Ia number density $\sigma^{2}_{obs}(r)/\bar{n}_{v}(r)$ for the velocity field. The average PV error is given in terms of a fractional distance error $\alpha$, and a contribution from random motions $\sigma_{obs,rand}=300\,\mathrm{kms^{-1}}$,
\begin{equation}
\sigma^{2}_{obs}(r) = (\alpha H_{0}r)^{2} + \sigma^{2}_{obs,rand}.
\label{eq:err}
\end{equation}
Finally, we model the relevant power spectra using
 \begin{align}
P_{\delta \delta}(z(r),k,\mu_{\phi}) &= \left(\frac{1}{\beta^{2}(z)}+\frac{2\mu_{\phi}^{2}}{\beta(z)}+ \mu_{\phi}^{4}\right)(f(z)\sigma_{8}(z))^{2} \nonumber \\ 
& D_{g}^{2}(k,\mu_{\phi})\frac{P_{mm}(k,z)}{\sigma^{2}_{8}(z)}, \label{eq:pkbeg} \\
P_{\delta v}(z(r),k,\mu_{\phi}) &= \frac{H(z)\mu_{\phi}}{k(1+z)}\left(\frac{1}{\beta(z)} + \mu_{\phi}^{2}\right)(f(z)\sigma_{8}(z))^{2} \nonumber \\ 
& D_{g}(k,\mu_{\phi})D_{u}(k)\frac{P_{m\theta}(k,z)}{\sigma^{2}_{8}(z)},
\end{align}
\begin{align}
P_{vv}(z(r),k,\mu_{\phi}) &= \frac{H^{2}(z)\mu_{\phi}^{2}}{k^{2}(1+z)^{2}}(f(z)\sigma_{8}(z))^{2} \nonumber \\
& D^{2}_{u}(k)\frac{P_{\theta \theta}(k,z)}{\sigma^{2}_{8}(z)}, \label{eq:pkend} \\
D_{g}(k,\mu_{\phi}) &= \left[1+\frac{(k\mu_{\phi}\sigma_{\delta})^{2}}{2}\right]^{-1/2} \mathrm{and} \\
D_{u}(k) &= \mathrm{sinc}(k\sigma_{v}).
\end{align}
We have written the above models in terms of the redshift corresponding to a given comoving distance $z(r)$ ($H(z)$ is the Hubble parameter at this redshift) and in a particular way to highlight the parameters of interest $\boldsymbol{\lambda}=\{f(z)\sigma_{8}(z)$, $\beta(z)$, $\sigma_{\delta}$, $\sigma_{v}\}$.  The power spectra ${P_{mm}(k,z)}$, ${P_{m\theta}(k,z)}$ and ${P_{\theta \theta}(k,z)}$ are the real-space matter and velocity divergence auto- and cross-power spectra for the dark matter field and are computed using the implementation of two-loop Renormalised Perturbation Theory \citep{Crocce2006} found in the {\sc copter} numerical package \citep{Carlson2009}.

The combination $f(z)\sigma_{8}(z)$ is the normalised growth rate that we present forecasts for in this work. We use this combination as both $f$ and $\sigma_{8}$ are degenerate on linear scales, however their combination can still be used to constrain gravitational models even without explicit knowledge of $\sigma_{8}$ \citep{Song2009} and is what is typically measured using RSD and PV surveys. $\beta(z)=f(z)/b(z)$ is the ratio of the growth rate over the galaxy bias and here is treated as one of the nuisance parameters we marginalise over. We also marginalise over two additional nuisance parameters, $\sigma_{\delta}$ and $\sigma_{v}$, which characterise the non-linear damping of the density and velocity fields due to RSD. These are used as inputs to Lorenztian (for the density field) and sinc (for the velocity field) functions which reduce the power spectra on small scales but leave them unchanged on large scales. For these parameters we adopt the same values as used in \cite{Howlett2017}, $\sigma_{\delta}=4.24\,h^{-1}\mathrm{Mpc}$ and $\sigma_{v}=13.0\,h^{-1}\mathrm{Mpc}$, which were found to reproduce the effects of non-linear RSD in simulations \citep{Koda2014}.

The redshift dependence of the normalised growth rate and bias is included using $f(z)\sigma_{8}(z)=\Omega_m(z)^{0.55}\sigma_{8}(z=0)D(z)$ and $b(z)=b(z=0)D^{-1}(z)$, with $D(z)$ given by Eq.~\ref{eq:growthfactor} for $a=1/(1+z)$ and normalised to unity at $z=0$. Computing the necessary non-linear real-space power spectra is slow, so the redshift dependence is captured by interpolating the power $P(k,z)$ at each $k$ from a set of precomputed power spectra in the range $z=[0.0, 0.5]$ with $\Delta z = 0.05$. We do not include any redshift dependence in $\sigma_{\delta}$ or $\sigma_{v}$.

We compute forecasts for the selections presented in Section~\ref{sec:LSST}; the volumetric rate with varying numbers of SNe~IA with measured light curves and the number of SNe~IA with host redshifts from Taipan, WALLABY and a $J<19.0$ survey. We compare these to the constraints using only RSD measured in the DESI-BG and 4MOST-BG surveys, i.e., where only the $P_{\delta \delta}(r,k,\mu_{\phi})$ element of $\boldsymbol{\mathsf{C}}(r,k,\mu_{\phi})$ is non-zero. For all surveys we assume a value for the galaxy bias $b(z=0)=1.34$ to allow for a simpler comparison between results. The sky area for the SNe~IA surveys is taken to be $18,000\,\mathrm{deg^{2}}$, whilst we use $15,000\,\mathrm{deg^{2}}$ for the RSD surveys, which closely matches the current design of 4MOST and DESI. For all SNe~Ia samples we consider distance errors of both $10\%$ ($\alpha=0.1$) and $5\%$.

We do not account for potential systematic errors in any of our forecasts, however a discussion of how SN~Ia systematics could affect measurements of the growth rate is given in Section~\ref{sec:systematics}.

\subsection{Results} \label{sec:results}

The percentage errors on the normalised growth rate, $f\sigma_{8}$ in bins of $\Delta z = 0.05$ between $z=0.0$ and $z=0.5$, and for the full redshift range, are listed in Table~\ref{tab:vel}. The volumetric rate forecasts listed are those for the two limiting cases of only SNe~IA we expect to have light curves measured with LSST and for all LSST-detected SNe~IA. For both of these we also give constraints from RSD only, i.e., the constraints using only the redshifts of the SNe~IA to measure the density-density power spectrum, neglecting the additional information from their light curves. This shows the relative improvement when SNe~IA PVs are added. For the SNe~IA samples with pre-existing redshifts (last two columns) we emphasise that the constraints are from a combination of all the measured redshifts for these samples plus the much smaller number of SNe~IA which add to the growth rate constraints from RSD alone.

\floattable
\begin{deluxetable}{lcccccccc}
\rotate
\tabletypesize{\small}
\tablewidth{0pt}
\tablecolumns{9}
\tablecaption{Forecasts for the percentage error on the normalised growth rate $f\sigma_{8}$ for the RSD-only 4MOST-BG and DESI-BG surveys and for samples containing LSST SNe~Ia.}
\tablehead{\colhead{\multirow{2}{*}{Redshift}} & \colhead{DESI-BGs\tablenotemark{a}} & \colhead{4MOST-BGs\tablenotemark{b}} & \multicolumn{2}{c}{All LSST-detected SNe~Ia\tablenotemark{c,d}} & \multicolumn{2}{c}{LSST light curves Only\tablenotemark{c,e}} & \colhead{Taipan+WALLABY+SN~Ia\tablenotemark{c,f}} & \colhead{$J<19.0$+SN~Ia\tablenotemark{c,g}} \\ & \colhead{RSD-only} & \colhead{RSD-only} & \colhead{RSD-only\tablenotemark{h}} & \colhead{RSD+PVs} & \colhead{RSD-only\tablenotemark{h}} & \colhead{RSD+PVs} & \colhead{RSD+PVs} & \colhead{RSD+PVs}}
\startdata
$0.00 < z < 0.05$ & 56.8 & 57.1 & 66.3 & 20.1 (13.9) & 106.6 & 41.0 (27.5) & 25.4 (16.6)  & 24.3 (15.7)\\
$0.05 < z < 0.10$ & 21.5 & 21.6 & 24.6 & 11.5 (7.3)  & 38.5  & 22.7 (14.6) & 15.9 (11.4)  & 14.6 (9.8) \\
$0.10 < z < 0.15$ & 13.2 & 13.2 & 14.8 & 9.0 (5.8)   & 22.6  & 16.6 (11.4) & 11.8 (10.4)  & 10.6 (8.3) \\
$0.15 < z < 0.20$ & 9.6  & 9.7  & 10.6 & 7.5 (5.0)   & 15.8  & 13.0 (9.5)  & 9.1 (8.6)    & 8.2 (6.9) \\
$0.20 < z < 0.25$ & 7.7  & 7.6  & 8.3  & 6.3 (4.4)   & 12.1  & 10.5 (8.2)  & 7.4 (7.2)    & 6.7 (6.0) \\
$0.25 < z < 0.30$ & 6.5  & 6.4  & 6.8  & 5.5 (4.0)   & 9.8   & 8.8 (7.1)   & 7.2 (7.1)    & 5.7 (5.3) \\
$0.30 < z < 0.35$ & 5.8  & 5.5  & 5.8  & 4.9 (3.7)   & 8.2   & 7.6 (6.4)   & 8.3 (8.3)    & 5.0 (4.8) \\
$0.35 < z < 0.40$ & 5.5  & 5.0  & 5.1  & 4.4 (3.4)   & 7.1   & 6.7 (5.7)   & 13.9 (13.9)  & 4.5 (4.4) \\
$0.40 < z < 0.45$ & 5.9  & 4.8  & 4.6  & 4.1 (3.2)   & 6.3   & 6.0 (5.2)   & -            & 4.2 (4.2) \\
$0.45 < z < 0.50$ & 10.9 & 5.8  & 4.2  & 3.8 (3.0)   & 5.7   & 5.4 (4.8)   & -            & 4.1 (4.0) \\
\textbf{0.00 $< \mathbf{z} <$ 0.50} & \textbf{2.5} & \textbf{2.2} & \textbf{2.1} & \textbf{1.8 (1.3)} & \textbf{2.9} & \textbf{2.7 (2.2)} & \textbf{3.4 (3.2)} & \textbf {1.9 (1.7)} \\ \tableline
\enddata
\tablenotetext{}{\textbf{Notes}.}
\tablenotetext{a}{Using number densities from \cite{DESI2016}. Redshifts for all galaxies, no PVs.}
\tablenotetext{b}{Using the number density of objects expected in the 4MOST-BG survey. Redshifts for all galaxies, no PVs.}
\tablenotetext{c}{Assuming $10\%$ ($5\%$) distance errors.}
\tablenotetext{d}{For all SNe~IA \textit{detected} by LSST as described in Section~\ref{sec:volrate}, assuming accurate light curves, redshifts and PVs (from LSST or otherwise) from every SN~IA.}
\tablenotetext{e}{Assuming redshifts and PVs from only the $\sim 50,000\,\mathrm{SN}$~$\mathrm{Ia\,yr^{-1}}$, detailed in Section~\ref{sec:volrate}, that could have accurate light curves from LSST alone.}
\tablenotetext{f}{For the Taipan+WALLABY target selection in Section~\ref{sec:taipan}. Redshifts for all TAIPAN+WALLABY galaxies. PVs from the $\sim 18,000$ SNe~IA found in those galaxies.}
\tablenotetext{g}{For the $J<19.0$ selection in Section~\ref{sec:futuresurvey}. Redshifts for all $J<19.0$ galaxies. PVs from the $\sim 160,000$ SNe~IA found in those galaxies.}
\tablenotetext{h}{Constraints when only the redshifts are used, regardless of available light curve measurements.}
\label{tab:vel}
\end{deluxetable}

We find similar constraints for the DESI-BG and 4MOST-BG surveys, reflecting their similar design and the fact that, as they only use RSD, these surveys quickly reach the cosmic variance limit at low redshift. The SNe~Ia PVs allow us to break this limit as they sample the same underlying structure as the RSD measurements. This is most apparent at the lowest redshifts, where the volumetric rate predictions show a factor of $\sim2$ improvement over the RSD constraints, and where the $J<19.0$ sample has significantly better constraints even though the selection function is similar to that of the 4MOST-BG sample. 

\begin{figure*}
\centering
\includegraphics[width=0.49\textwidth]{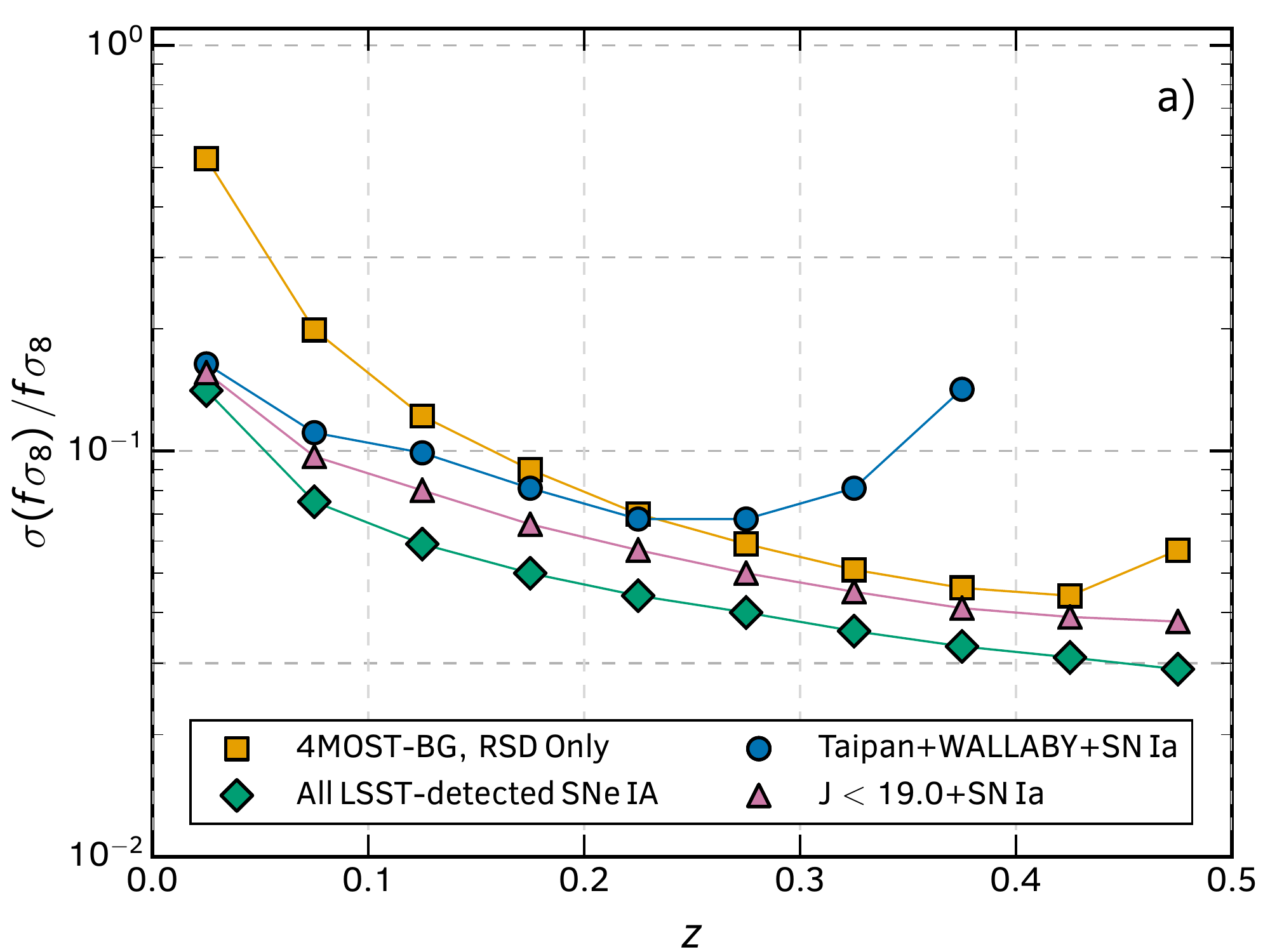}
\includegraphics[width=0.49\textwidth]{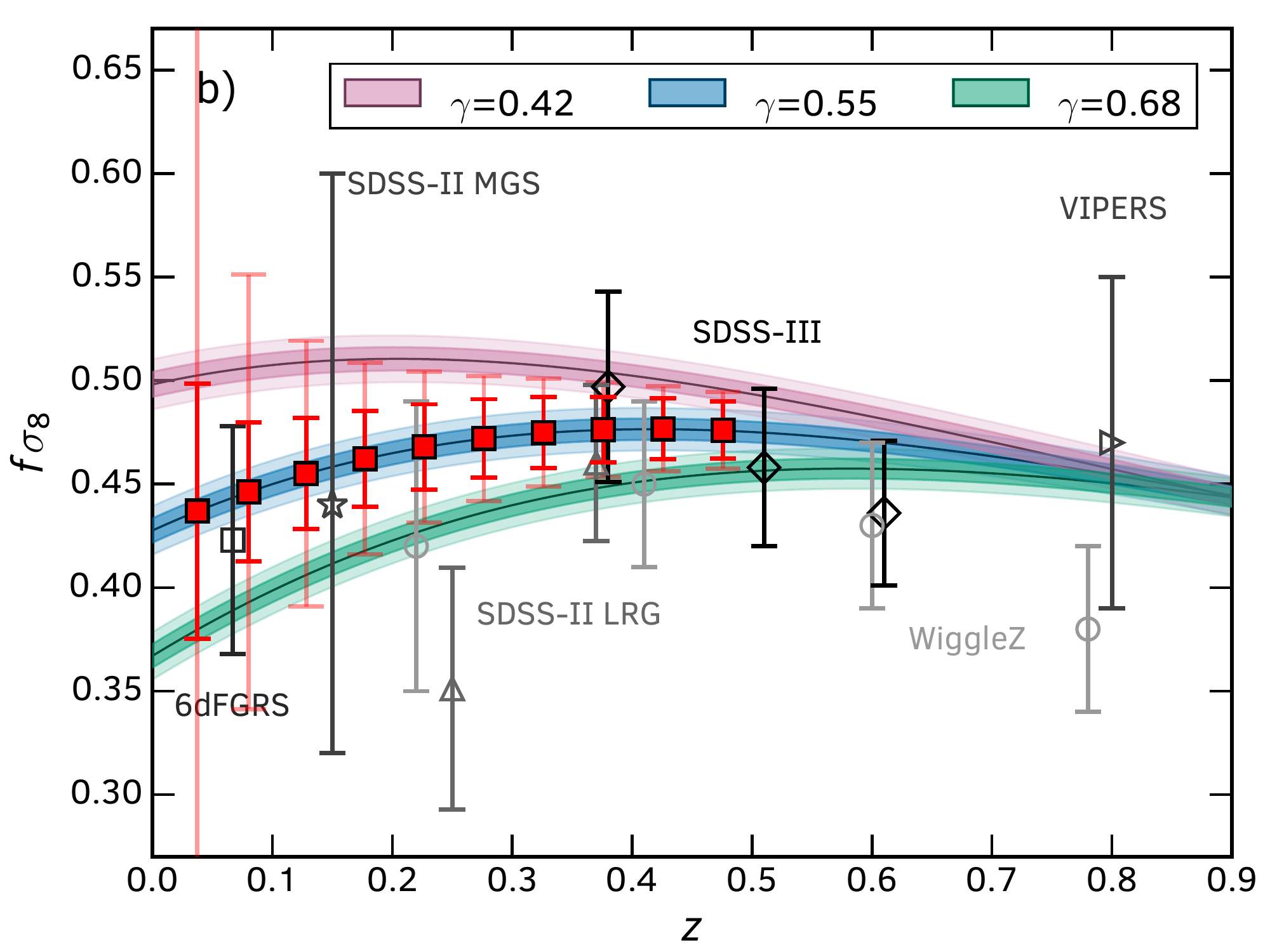}
  \caption{Forecasts for our three SNe~Ia samples assuming $5\%$ distance errors. \textbf{a).} Fractional errors as a function of redshift plotted against the 4MOST-BG sample. \textbf{b).} A comparison of the volumetric rate forecasts for all LSST-detected SNe~IA (red squares) against existing measurements \citep{Beutler2012,Howlett2015,Oka2014,Alam2016,Blake2011a,Blake2011b,delaTorre2013} and predictions from Planck (\citealt{Planck2016}; normalised at the redshift of recombination) with different values for $\gamma$. These $f\sigma_{8}$ predictions as a function of redshift are calculated self-consistently using the method in \citealt{Howlett2015} (Eqs.~26-30 therein), which accounts for the fact that the growth factor \textit{cannot} be evaluated from Eq.~\ref{eq:growthfactor} for different values of $\gamma$. The outer error-bars for the volumetric rate measurements show RSD-only constraints (using only the redshift measurements of the SNe~IA and neglecting their light curves); the inner show those including SNe~Ia PVs. This highlights the redshift-dependent improvement due to the SNe~Ia PVs.
  }
  \label{fig:growthrate}
\end{figure*}

The fractional errors for the RSD-only 4MOST-BG sample, all LSST-detected SNe~Ia and our two samples where we only use SNe~IA that are likely to already have host redshifts are plotted in Fig.~\ref{fig:growthrate}. The right-hand panel of this Figure then compares the LSST-detected SNe~Ia constraints with and without SNe~IA PVs against current measurements and the predictions from different models of gravity. For SNe~Ia that are likely to already have host redshifts, the Taipan+WALLABY+SN~Ia sample achieves better constraints than 4MOST or DESI below $z\approx0.15$, but at higher redshifts the number of galaxies drops significantly resulting in poor constraining power. For the $J<19.0$ sample the constraints are again comparable or better than with RSD-only for all redshift bins. This is because at low redshift the SNe~Ia provide an increase in constraining power, whilst at high redshift we still obtain large numbers of galaxies and can constrain the growth rate via RSD, using the SNe-Ia to break the degeneracy with any nuisance parameters.

\begin{figure*}
\centering
\includegraphics[width=0.98\textwidth]{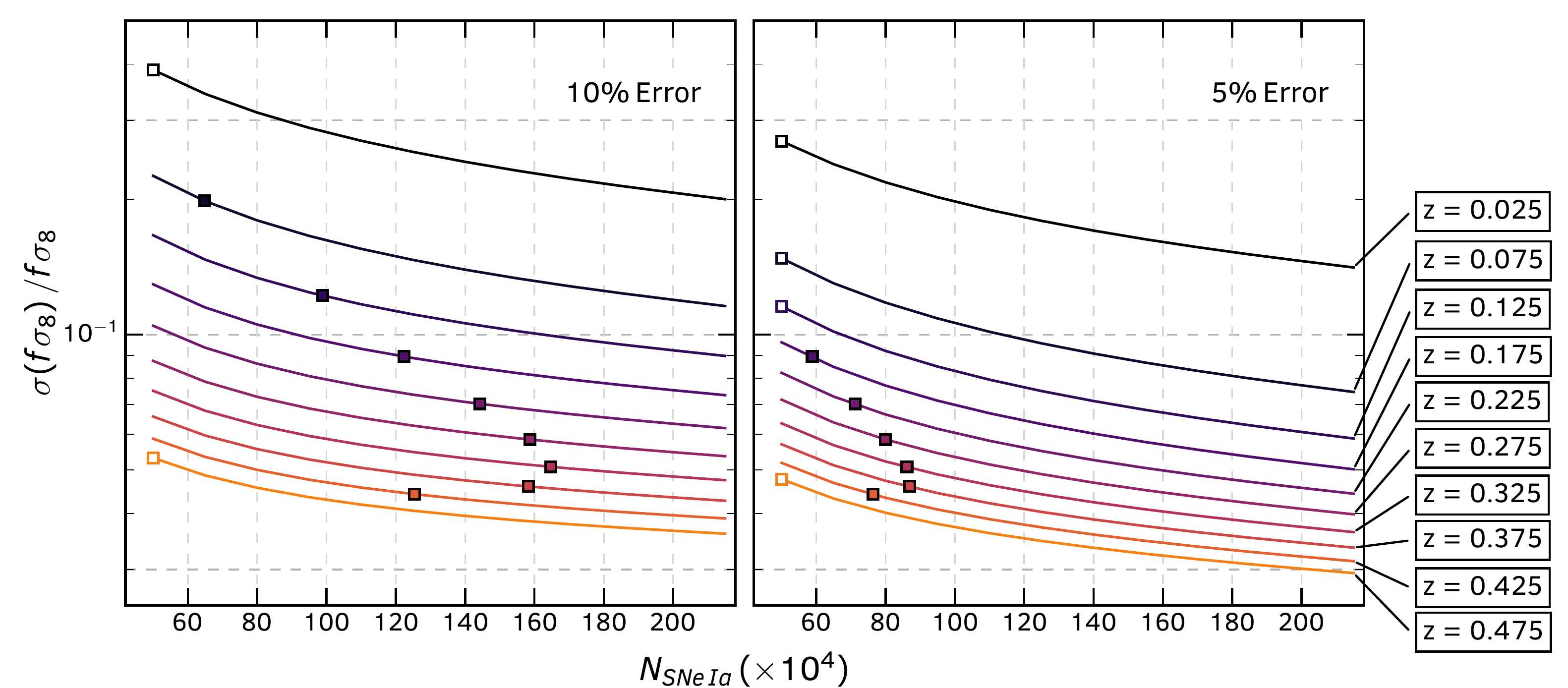}
  \caption{Forecasts for the fractional error on the growth rate as a function of SNe~Ia with precise distance determinations, assuming distance errors of 10\% (left) and 5\% (right). Although survey-independent, the lower limit of the x-axis is equivalent to the SNe~IA we expect to have light curves measured by LSST itself, whilst the upper limit is the expected number detected by LSST (for which distances measurements would require additional follow-up). Different lines represent different redshift bins of width $\Delta z = 0.05$. Points represent the intercept of each line with the RSD-only forecasts for the 4MOST-BG survey, and so allow us to infer the number of distance measurements necessary to improve over the 4MOST-BG constraints in each bin. Finally, open symbols represent cases where the light curves we expect from LSST alone already have greater predicted constraining power than 4MOST.
  }
  \label{fig:growthratensn}
\end{figure*}

In Figure~\ref{fig:growthratensn}, we demonstrate how the growth rate predictions change as we increase the number of SNe~IA with distance measurements beyond those we are likely to obtain with LSST alone. We also plot the intersect of the fractional error as a function of the number of SNe~IA in each redshift bin with the prediction using RSD from the 4MOST-BG sample. This intersect point highlights how many SNe~IA with measured distances we would require in each redshift bin to improve over the constraint from 4MOST using RSD. 

We find that, assuming $5\%$ distance errors, the SNe~IA distances we could obtain with LSST are already sufficient to improve over the RSD constraints below $z=0.15$ and that measuring light curves to similar accuracy for only a modest fraction of the remaining LSST-detected SNe~IA allows for superior constraints across the full redshift range we consider. For $10\%$ distance errors, the required number of SNe~IA is larger, but we can still improve on the 4MOST-BG constraints for all redshift bins using some fraction of the total number of SNe~IA LSST will detect. We emphasise here that, unlike our $J<19.0$ predictions, the total number of objects (with SNe~IA light curves and redshifts) even for all LSST detections is a factor of $\sim 4$ less than the number of galaxies in the 4MOST or DESI BG samples, which demonstrates the superior constraining power of the peculiar velocity measurements.

We do not consider forecasts beyond $z=0.5$ as at higher redshift the SNe~Ia distance errors become large and the majority of the growth rate information comes from RSD (see Fig.~\ref{fig:growthrate}). Whilst at $z=0.5$ SNe~Ia still help in marginalising over the nuisance parameters, the constraining power of DESI and 4MOST improves significantly beyond this due to the large cosmological volumes they can probe with their Luminous Red Galaxy and Emission Line Galaxy samples. Combined, these can also be used to break the cosmic variance limit in the same way as a sample of SNe~Ia. Hence the SNe~Ia samples quickly become less competitive.

\section{Systematics} \label{sec:systematics}
In our analysis we have assumed spectroscopic classification of the supernovae and have not included SN~Ia systematics such as flux calibration or extinction correction errors, misclassification or the incorrect assignment of SN~Ia to their host galaxies. The ability of photometric estimators to classify supernovae given LSST quality data and the magnitude of any systematic effects expected within LSST is currently under investigation but has not been clearly defined and so has not been included quantitatively in the forecasts we have presented. Here we provide a qualitative discussion of the need for spectroscopic classification and how we expect different systematic effects to manifest in measurements of the growth rate using the two point correlations of the SNe~IA PVs. A more rigorous analysis, measuring the velocity power spectrum using simulations of SNe IA as detected by LSST and including such systematic effects is left for future work.

\subsection{SNe~IA measurement systematics}

Systematic errors within the flux calibration or extinction correction for a sample of SNe~IA can be described via a limiting systematic `error floor' in each redshift bin or across the full SNe~Ia sample \citep{Linder2003,Kim2011}, such that for large numbers of SNe~IA, the error on the mean distance measured in a given redshift bin does not continue to decrease purely in proportion to the square root of number of SNe~IA. A systematic offset in the distance modulus resulting from this systematic error would act as an error on the zero-point of the SNe~IA PVs, which is also present with other methods used to estimate PVs such as the TF and FP relations. 

This is an issue for measurements of the bulk flow, where the zero-point acts in the same way as the bulk motion of the local universe, and a systematic error can bias bulk flow constraints. However, the velocity power spectrum is sensitive to the \textit{variance} of the PVs as opposed to the mean, which is limited by the intrinsic dispersion in the distance indicator. In this way \cite{Howlett2017} showed that errors in the zero-point simply act as additional shot-noise in the velocity power spectrum and as long as the systematic errors are small compared to the intrinsic scatter, the effect of this on growth rate constraints is negligible. Alternatively, the additional shot-noise component can be marginalised over analytically and at little cost to the growth rate constraints \citep{Johnson2014, Howlett2017b}. In terms of quantities relevant to SNe~IA, \cite{Linder2003,Kim2011} consider a systematic error in the distance modulus of around $\sigma_{sys}\approx 0.03$ with some dependence on redshift. This is significantly less than even the lowest intrinsic dispersion we use in this work and might expect for future SNe~IA, $\sigma_{int}\approx 0.1$ and so we expect that the precision of the growth rate forecasts presented here will be unaffected by the inclusion of this systematic error.

\subsection{Photometric Classification and associated systematics}

Photometric classification of supernovae is an active area of study, with high-redshift supernova searches such as those in the Dark Energy Survey \citep{DES2005} and LSST planning to perform photometric classification to define their cosmological samples (i.e., \citealt{Campbell2013}). However, current photometric classifiers, either template-based (e.g. \citealt{Sako2011}) or using machine learning (e.g. \citealt{Lochner2016}) are not perfect and introduce both systematic errors and potential biases into cosmological studies. A particularly subtle problem is that the very features useful for photometric classification: flux, color, light curve shape etc., are the same statistics used to determine supernova distances.  This leads to strong covariance between an objects classification and distance measurement, whose impact in cosmological studies has yet to be studied in full. As the effects of these on LSST-quality data and cosmological analyses in general is still not well understood, we have assumed for simplicity that we get spectroscopic redshifts of the host galaxies and classification of the supernova itself in our forecasts. This also negates the effects of SNe misclassification and host misidentifications. In the absence of spectroscopic classification, we would expect systematic errors due to both of these and discuss their expected impact on our forecasts below. Overall, the requirements for spectroscopic follow-up for measuring accurate SN Ia peculiar velocities may be relaxed, depending on progress in photometric classification over the coming years.

\subsubsection{Misclassification}

Misclassification of supernovae as SNe~IA leads to contamination in the sample and incorrect distance inference. Photometric estimators typically also miss some fraction of true SNe~IA. Accounting for completeness or false positives in the photometric classification of the SNe~IA in our forecasts would reduce the total number of usable SNe~IA, increase the shot-noise in our measurements of the velocity power spectrum and reduce the constraints on the growth rate. However, the factor of 0.4 we have used in this work as the fraction of SNe~IA LSST will detect pre-peak luminosity already carries considerable uncertainty, such that the effects of completeness on our growth rate forecasts are likely small compared to the current uncertainty in the factor of usable SNe~IA we have assumed. Furthermore, we have provided forecasts assuming distance errors of both $5\%$ and $10\%$, which can include contributions from both statistical and systematic errors. Even with the effects of misclassification of SNe, we consider distance errors of $~10\%$ to be conservative. Finally, it is worth noting that Type II-P SNe also show promise as `standardizable' candles \citep{dAndrea2010,deJaeger2017} in the $z<0.5$ universe, and are expected to be detected in even greater numbers with LSST than SNe~IA \citep{Ivezic2008}. PVs from such a sample have the potential to significantly improve over the forecasts presented here for SNe~IA alone, even accounting for completeness and systematic errors.

\subsubsection{Host Misidentification}

While spectroscopic classification also provides a supernova redshift whose consistency can be tested with that of the purported host, the lack of that consistency test leads to misidentification of the host galaxy \citep{Gupta2016}. In the event that the true and assumed host galaxy are physically close, this is not an issue for PV measurements. In fact, a common practice is to use group galaxy catalogues measured from redshift surveys (i.e., \citealt{Crook2007}) to assign identical redshifts to PV targets belonging to the same group, which partially removes the effects of non-linear motion on the measured PVs \citep{Hong2014,Springob2014}. In this sense, the SNe~IA would be given the same observed redshift regardless of the host it is assigned to. 

In the case of incorrect assignment of SNe~IA to host galaxies that are close in angular separation but physically far apart, we expect to be able to remove these after the PVs have been measured. On linear scales the peculiar velocities (excluding statistical errors) are expected to be Gaussian distributed. Hence, for physically distinct galaxies, the difference between the redshift distance and the true distance SNe~IA measurements is likely to lead to an abnormally large PV, which can be then be removed via sigma-clipping, as was done for TF-based PVs in \cite{Howlett2017b}.

\section{Conclusions} \label{sec:conclusion}

We have demonstrated that LSST SNe~Ia could provide measurements of the $z<0.5$ growth rate that are more precise than those available using only RSD from DESI or 4MOST. Our best constraints come from the case where we are able to obtain host redshifts, lightcurves and spectroscopic classification for all $\sim 2.2\times10^{6}$ LSST-detected SNe~Ia, based on the volumetric SNe~Ia rate from \cite{Dilday2010}. There is currently no planned survey that can accomplish this; LSST is expected to obtain sufficiently accurate light curves for at most $\sim 500,000$ of these, however the target density, $\sim12\,\mathrm{deg^{-2}\,yr^{-1}}$, is small and could be accommodated as part of a larger survey programme. We have also relaxed this condition and looked at how many SNe~IA would be required to achieve constraints comparable to those from DESI or 4MOST at various redshifts, finding that SNe~IA with lightcurves measured from LSST \textit{alone} could do better than RSD below $z=0.15$, given accurate classification and host redshifts. We expect many of these local SNe~IA to already have host redshifts from upcoming galaxy redshift surveys.

To further explore this, we have combined simulated galaxy catalogues with a prescription for the SNe~Ia rate as a function of stellar mass and SFR and explored those SNe~Ia that could already have host redshifts from upcoming large galaxy surveys. Our test cases include Taipan, WALLABY, and a future multi-object spectroscopic survey. We find that a $J<19.0$ magnitude-limited sample could obtain $\sim160,000$ host redshifts. Although the number of SNe~Ia is much smaller than the volumetric rate, predictions for the growth rate from this sample still outperforms those using only RSD with DESI or 4MOST. Hence, variations of the 4MOST or DESI target selections could allow for a large number of host redshifts that can be used to significantly augment and improve the constraining power of these surveys.

In this work, our primary aim is to motivate further consideration of the potential of LSST detected SNe~IA to measure the growth rate and test gravity. As such we have assumed spectroscopic classification for our SNe and ignored potential systematic effects. This is also partly driven by our limited current understanding of both the ability of LSST-quality photometry to overcome these effects and the covariance and bias introduced into measurements of SNe~IA distances when using photometric classification methods. We have discussed how we expect various measurement systematics to manifest in measurements of the growth rate and anticipate that the assumption of spectroscopic classification can be relaxed as photometric estimators progress. Future studies will allow us to quantify the effects of various systematics and classification algorithms on the velocity power spectrum we will measure with LSST SNe~IA, and this work motivates a careful study of these in the context of testing gravity.

\acknowledgements
We thank Bob Nichol, Eric Linder, Patrick McDonald, David Parkinson and Chris Blake for their comments and the latter for providing the number density of the 4MOST-BG sample. This research was conducted by the Australian Research Council Centre of Excellence for All-sky Astrophysics (CAASTRO), through project number CE110001020. This research has made use of NASA's Astrophysics Data System Bibliographic Services and the \texttt{astro-ph} pre-print archive at \url{https://arxiv.org/}. All plots in this paper were made using {\sc matplotlib} \citep{Hunter2007}.

\end{document}